# A complex network theory approach for optimizing contamination warning sensor location in water distribution networks

**Rezvan Nazempour [a], Mohammad Ali Saniee Monfared [b1], Enrico Zio [c2]**

[a] Department of Industrial Engineering, School of Engineering, Alzahra University, Tehran, Iran

[b] Department of Industrial Engineering, School of Engineering, Alzahra University, Tehran, Iran

[c] Energy Department, Politecnico di Milano, Via Ponzio 34/3, 20133 Milano, Italy

**Abstract**

Drinking water for human health and well-being is crucial. Accidental and intentional water contamination can pose great danger to consumers. Optimal design of a system that can quickly detect the presence of contamination in a water distribution network is very challenging for technical and operational reasons. However, on the one hand improvement in chemical and biological sensor technology has created the possibility of designing efficient contamination detection systems. On the other hand, methods and tools from complex network theory, which was primarily the domain of mathematicians and physicists, provide analytical output for engineers to design, optimize, operate, and maintain complex network systems such as power grids, water distribution networks, telecommunication systems, internet, roads, supply chains, traffic and transportation systems.

In this work, we develop a new modeling approach for the optimal placement of sensors for contamination detection in a water distribution network. The approach originally combines classical optimization and complex systems theory.

## 1. Introduction

[1] Corresponding author: Mohammad Ali Saniee Monfared, School of Engineering, Alzahra University, Vanak Street, P.O. Box 1993891167, Tehran, Iran. Email: mas_monfared@alzahra.ac.ir;mas_monfared@yahoo.com

[2] enrico.zio@ecp.fr, Chair on Systems Science and Energetic Challenge Fondation EDF (Electricite' de France) CentraleSupelec , and Rector's delegate and President of the Alumni Association of the Politecnico di Milano, Italy Rector's delegate for Individual Fundraising.

Drinking water is a physical, cultural, social, political and economic resource, crucial for human health and well-being. Safe supply of potable water is a basic necessity of mankind in the industrialized society; therefore, water supply systems are a most important public utility [1].Water Distribution Network (WDN) is a network of interconnected pipes and other appliances through which water is conveyed, stored intermittently and pumped where necessary in order to meet the demand and pressure requirements of the system. A WDN is a large, spatially distributed, complex infrastructure, where the possibility of human interference is high. In general, water distribution systems can be divided into four main components: (1) water sources and intake works, (2) treatment works and storage, (3) transmission mains, and (4) distribution network [1]. Accidental and intentional water contamination can be of great danger to consumers. Spread of contaminated water in Milwaukee, US, in 1993 affected more than 400,000 people, when a microorganism was transported through the distribution system [2]. An outbreak of water-borne disease epidemics in Walkerton, Ontario, Canada, in the year 2000 affected 2,300 people, as a result of exposure to contaminated drinking water [3]. The challenge to designing a monitoring system of a limited number of sensors to detect accidental or intentional contaminant intrusion events in a WDN has attracted significant attention in recent years. Rathi and Gupta in [4] classified the methodologies for sensor placement into two categories, addressing single objective and multi-objective sensor location problems. The models and methods proposed by different researches to solve these problems are quite varied [5].

A WDN is a complex network. Complex networks are defined as systems whose structure is irregular, distributed, complex and dynamically evolving in time. In such networks, structure affects function [6]. Historically, the study of complex networks has been mainly the domain of graph theory, which was introduced by Leonhard Euler in 1736 [7]. Random graph [8], Small World [9] and, Scale Free [10] models have been developed to describe different network structures.

In the last decades, the study of complex networks has grown rapidly and has attracted many researchers from different academic disciplines and application fields like, for example, in power grids [11]–[14], transportation networks [15]–[18], and water distribution networks [19]–[21].

The aim of the paper is to present a new modeling approach for the optimal placement of sensors for contamination warning in a WDN. This problem has been investigated by the operational research community as a single and bi-objective optimization problem [22]–[27]. Also design, operation and maintenance of WDNs have recently been investigated from a complex network theory perspective [28]–[31]. In this paper we combine classical optimization and complex network theory in an innovative way.

The rest of this paper is organized as follows. Section 2 presents some minimum background on complex network theory. Section 3 presents the work in terms of assumptions and scenario planning, problem formulation and solution method. In Section 4, we introduce a well-known WDN as case study. Then, we present the results obtained by applying the proposed optimization model to the case study WDN in Section 5. Finally, conclusions and future research steps are drawn in Section 6.

## 2. Complex network

There are many complex networks in nature, society and technology which are comprised of multiple components joined by various types of interconnections and links: for example, social networks consisting of people and their contacts, the World Wide Web as an information network consisting of the web pages and links among them, and technological networks such as the Internet comprised of the computers and physical connections among them [32]. Design, operation and maintenance of WDNs have recently been explored by few researchers using the complex network approach.

Yazdani and Jeffrey [33], for the first time, represented a WDN as a large sparse planar graph with complex network characteristics and evaluated the topological features and resilience of the network. They proposed two metrics of meshed-ness and algebraic connectivity for quantification of redundancy and robustness, respectively, which could be used in optimization design models [28]. They also used empirical data from benchmark networks to study the interplay between network structure and operational efficiency, reliability, and robustness [34]. Further, they modeled WDNs as weighted and directed graphs by using the physical and hydraulic attributes of system components, and proposed a new measure of component criticality, the demand-adjusted entropic degree, to support identification of critical nodes and their ranking according to failure impacts [29]. Yazdani et al. [19], also proposed four strategies (branched, looped, extra-looped and perfect-mesh) for WDN expansion aimed at securing and promoting structural invulnerability subject to design and budget constraints, considering a developing country case study.

Gutiérrez-Pérez et al. [30], [35], introduced spectral measures to establish vulnerability areas in WDNs and proposed an index for generating a topological importance measure to form clusters in a WDN. Hawick [36], applied graph theory to some synthetic and real WDNs and studied the robustness and fragmentation properties through simulated component failure. Sheng et al. [31], explored the formation of isolated communities in WDNs based on complex network theory. Shuang et al. [20], evaluated the nodal vulnerability of WDNs under cascading failures.

### 2.1. Characterization of a complex network

A complex network is often represented as a directed (or undirected), weighted (or unweighted) graph $G = (N, L)$, with $N$ a set of nodes representing the dynamical units and $L$ a set of edges representing the interactions between them. A binary graph can be completely described by its adjacency (or connectivity) matrix $A$, an $N \times N$ square matrix whose entry $a_{ij}$ ($i, j = 1, \ldots, N$) is equal to 1 when the link $l_{ij}$ exists, and 0, otherwise [14].

A graph of a complex network is characterized and quantified by some metrics measuring the topological and dynamical properties of the network. Such metrics are important in understanding the network complexity and are instrumentals for the design, operation and maintenance of complex networks in practice. Here, we list eight metrics in Table 1 and illustrate their use for analyzing the graphs of five WDNs.

Table 1. Basic network metrics [13], [14]

| Metrics | Description | Formula |
|---|---|---|
| Maximum degree | Degree $k_i$ is defined for node $i$ as the number of links connected to the node. Then, $k_{max}$ is the maximum number of connections to a node in a network. | $k_{max} = \max\{k_i, \text{ for } i \in N\}$ |
| Mean degree | Average degree of all nodes in a network. | $<k> = \sum_{i=1}^{N} k_i / N$ |
| Mean shortest path length | Average of all shortest paths in a network. $N$ is the size of the network and $d_{ij}$ is the length of the shortest path from node $i$ to node $j$. | $<l> = \frac{1}{N(N-1)} \sum_{i,j \in N, i \neq j} d_{ij}$ |
| Clustering coefficient | The quantity $c_i$ is the local clustering coefficient of node $i$, expressing how the neighbors of two adjacent nodes have a link in between. Here, $e_i$ is the actual number of and $C$ is the average clustering coefficient. | $c_i = \frac{2e_i}{k_i(k_i-1)} = \frac{\sum_{jm} a_{ij} a_{jm} a_{mi}}{k_i(k_i-1)}$ $C = \frac{1}{N} \sum_i c_i$ |
| Diameter | Graph diameter is the maximum value of shortest paths between all nodes. | $d = \max\{d_{ij}\} \ \forall i,j \in G$ |
| Critical fraction | The fraction of nodes to be eliminated to render a graph unable to percolate, i.e., to go from one node to another following an edge. | $f_c$ |
| Betweenness centrality | A node is central if it structurally lies between many other nodes, in the sense that it is traversed by many of the shortest paths connecting pairs of nodes. Here, $n_{jk}$ is the | $b_i = \frac{1}{(N-1)(N-2)}$ |

| | number of shortest paths from node $k$ to node $j$, $n_{jk}(i)$ is the number of shortest paths that contain $i$ and $0 \leq b_i \leq 1$. | $\sum_{j,k \in G, j \neq k \neq i} \frac{n_{jk}(i)}{n_{jk}}$ |
|---|---|---|
| Closeness centrality | The central nodes are those which on average need fewer steps to communicate with the others (and not just the first neighbors). Here, $d_{ij}$ is the shortest path from node $i$ to node $j$ and $0 \leq cl_i \leq 1$. | $cl_i = \frac{N-1}{\sum_{j \in G} d_{ij}}$ |

In Figures 1 and 3, five different WDNs are shown. In Figure 1, we show WDNs for EastMersea, Colorado Springs, Richmond and Kumasi [37], and in Figure 3 we show the fifth WDN, i.e., the BWSN which we consider in details in Section 4 where we discuss our case study.

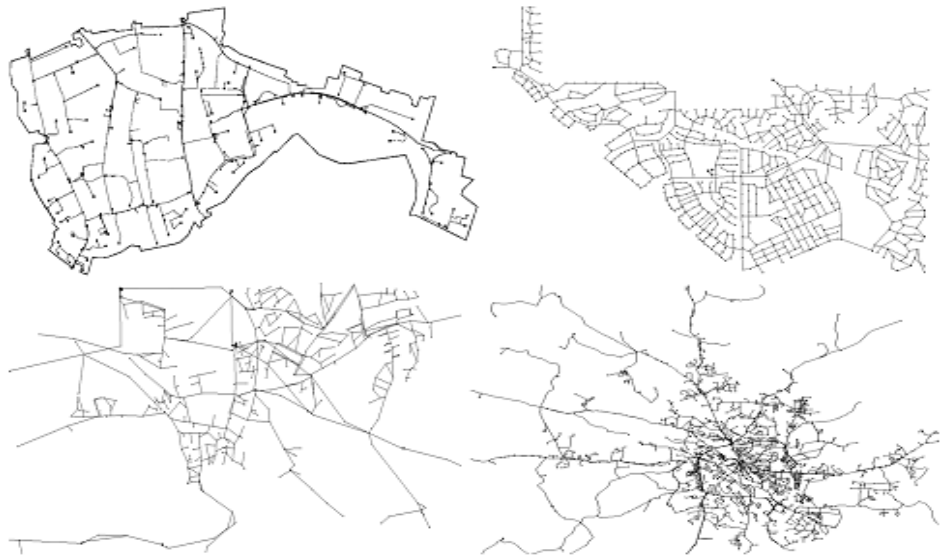

Figure 1. Four sample WDNs [37] EastMersea (up left), Colorado Springs (up right), Richmond (down left), Kumasi (down right)

Table 2 reports the values of the numerical metrics for these five WDNs. From these results, it become clear that some metrics are not changing from one WDN to another, e.g., $K_{max}$, $<k>$, because of the physical limitations of WDNs. For example, in any node or junction of a WDN we will not have more than 4 connections. Hence, such metrics are not useful in differentiating between different WDNs. On the contrary, other metrics in Table 2 vary from one network to another and, hence, can be used to differentiate between WDNs. For example, we use the values of $k_i$, $b_i$ and $cl_i$ that represent node degree, node betweenness and node closeness, respectively, to find out the most important nodes in the network and place sensors accordingly, as will be shown later (Figure 6). Note that in Table 2 we have reported the mean values for $k_i$, $b_i$ and $cl_i$ as $<k>$, $b$ and $cl$, for the five metrics considered.

We now move on to Section 3, where we develop our new modeling approach by which complex network theory and classical optimization theory are integrated to solve sensor placement problems in WDNs.

Table 2. Complex network properties of five WDNs

| Network | N | L | $K_{max}$ | $<k>$ | L | $C_b$ | d | $f_c$ | b | Cl |
|---|---|---|---|---|---|---|---|---|---|---|
| East-Mersea | 755 | 769 | 4 | 2.04 | 34.48 | 0.36 | 97 | 0.22 | 0.00019 | 0.039 |
| Colorado Springs | 1786 | 1994 | 4 | 2.23 | 25.94 | 0.42 | 69 | 0.42 | 0.00024 | 0.028 |
| Kumasi | 2799 | 3065 | 4 | 2.19 | 33.89 | 0.45 | 120 | 0.37 | 0.00009 | 0.009 |
| Richmond | 872 | 957 | 4 | 2.19 | 51.44 | 0.56 | 135 | 0.32 | 0.00006 | 0.072 |
| BWSN | 129 | 178 | 4 | 2.54 | 10.15 | 0.85 | 25 | 0.42 | 0.00750 | 0.041 |

## 3. Methodology

Selecting sensor locations for minimizing the impact from a set of plausible contamination incidents is a problem conceptually equivalent to the well-known problem of facility location [38]. This can be framed as the *p* facility location problem, in which *p* facilities must be located in such a way that the distance from each facility to its customers is minimized. The classic *p* facility location problem can formally be stated as follows. Consider the layout of a city, and imagine that *p* fire stations must be located in order to best serve the city's residents and infrastructure. Each house and building in the city is a customer, and each fire station a facility. Given a proposed set of locations, the problem objective is to minimize the average distance from a customer to the nearest fire station facility [39].

For the drinking water sensor placement problem, the sensors are considered as facilities like fire stations in the example above. Each contamination incident is a single "customer." A contamination incident propagates contaminated water through the network and is "served" from the network users' point of view, by the first sensor facility that detects the contamination.

The physical structure of a water distribution system is a network in which nodes represent water sources, tanks, and junctions (the connection between pipes and points of water withdrawal) and edges represent pipes, valves, and pumps. Contamination events are assumed to result only from contaminants introduced at system nodes. We also assume that the hydraulic behavior of a water distribution system is not substantially changed by the intentional injection of a contaminant [39] and that sensors are perfect (i.e., fully reliable), meaning that they can detect contaminants of any concentration with no false negatives and false positives.

Water flows from sources to consumers, possibly transporting contaminants if present in the water. Sensors can only detect contamination events that occur upstream from their location. Given a budget-constrained number of sensors, it is generally impossible to detect every possible contamination event in a water distribution network. Alternatively seen, we wish to identify as many of the contamination events as possible, with the budget-limited number of sensors available. Without knowing the distribution of contamination events across a water distribution network, we generally assume that every node has an equal chance of acting as the contamination origin [39].

Scenario planning is often used to represent the uncertainty of the future in business and governmental policy analysis, and in technological evaluation and development [40]. In our research, we use scenario planning to structure the uncertainty inherent in a contamination event. A contamination event will occur as one scenario from a large set of scenarios. The set of scenarios considered should be representative of the possible realizations of uncertain parameters in the problem.

A set of scenarios is represented as $\{(t, d, N) \mid t \in T_p; d \in D; N = N\}$, with each scenario starting from time $t$, at any of the $N$ nodes and, lasting a duration $d$, where $T_p$ is the set of discrete event times, $D$ is the set of contamination durations and $N$ is the node set. The starting time could be any within a chosen period. The duration of the contamination event could range from a few seconds to several days, or even longer if not detected timely. The upper limit of the event duration is set to be the required detection time. For instance, if a contamination event is required to be detected within two hours from its onset, then the longest contamination duration considered is two hours. For the accidental contamination event, in absence of additional knowledge it is assumed that contamination events may occur at any node with equal probability [39].

### 3.1. Problem formulation

In a conventional design of WDN the following mathematical model is often used [41]:

$$MinTc = \sum_{p \in P}(D_p, L_p)$$

st.

$$\sum_{i \in N} Q_{ni} - \sum_{j \in N} Q_{nj} = D_n - S_n, \quad \forall n \in N \quad (1)$$

$$\sum_{p \in l} \Delta H_p = 0, \quad \forall l \in L$$

$$H_n \geq H_n^{\min}, \quad \forall n \in N$$

In which $TC$ is the total cost, $D_p$ is the diameter of pipe $p$, $L_p$ is the length of pipe $p$, $Q_{ni}$ is the inflow into node $n$, $Q_{nj}$ is the outflow from node $n$, $D_n$ is the water demand on node $n$, $S_n$ is the reserved capacity at node $n$, $\Delta H_p$ is the head loss on pipe $p$, $H_n$ is the head on node $n$, $H_n^{\min}$ is the minimum head required on the node, $P$ is the total number of pipes in the network, $N$ is the total number of nodes and $L$ the is total number of loops in the network. In this model, the first constraint set is the law of mass conservation as shown in Figure 2 (right), the second set of constraints are the law of energy conservation as shown in Figure 2 (left), and the third set of constraints define the minimum head requirements.

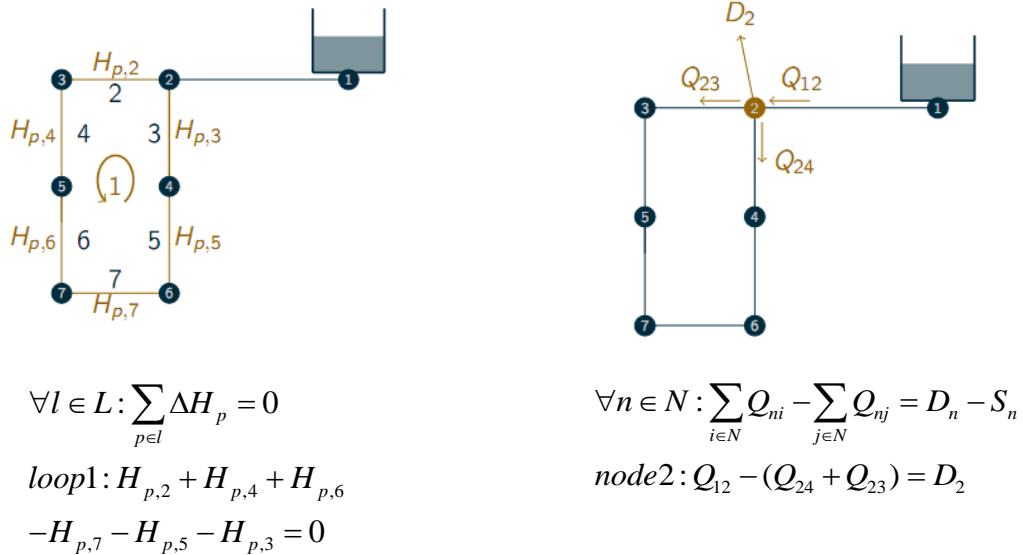

$$\forall l \in L : \sum_{p \in l} \Delta H_p = 0$$

$$loop1: H_{p,2} + H_{p,4} + H_{p,6} - H_{p,7} - H_{p,5} - H_{p,3} = 0$$

$$\forall n \in N : \sum_{i \in N} Q_{ni} - \sum_{j \in N} Q_{nj} = D_n - S_n$$

$$node2: Q_{12} - (Q_{24} + Q_{23}) = D_2$$

Figure 2. Conservation laws of mass (right) and energy (left) [41]

Now, we have to turn the hydraulic-based design optimization problem of a WDN presented above into a sensor placement optimization problem. This is the way for the classical approach to solving sensor placement optimization problems. Rathi and Gupta in [4] classified the methodologies on sensor placement into two categories, with reference to single-objective and multi-objective sensor location problems. The objectives function considered were: (1) early detection of contamination, e.g. minimum (or expected) time to detection (TD) [22], [42], [43], (2) minimizing the impact or consequences of contamination, e.g. minimum (or expected) volume of water consumed (VC) [23], [44], population exposed to contamination (PE) [24], [45], [46], and extent of contamination (EC) [25], [47], (3) covering as much as possible the population with limited number of sensors considering (i) minimizing risk (Risk) [48] and/or (ii) maximizing detection likelihood (DL) of contamination events [26], (4) minimization of sensor response time (SRT) [27], (5) minimization of the number of failed detections (NFD)/probability of failed detection (PFD) [49] and sensor detection redundancy (SDR) [50], [51]. Ostfeld et al. [5] compared the solutions provided by several algorithms based on four objectives: TD, PE, VC and DL, and concluded that the solutions provided by different algorithms were quite varied.

In this paper, we explore a new perspective by presenting an innovative bi-objective optimization model which combines classic optimization theory and complex network theory to solve a sensor placement problem. This modeling approach is illustrated in the following.

### 3.1.1. Objective functions

As mentioned earlier, different objective functions have been adopted in the current literature on sensor location optimization. Here, we adopt the widely used concept of coverage in the WDN as our first objective function $F_1$: node "$i$" in the WDN is covered if a contamination event at node $i$ can be detected by any of the sensors within time $T$, after the event has occurred. We refer to time $T$ as the ''coverage time''. For coverage maximization, we actually minimize the number of nodes which are not covered in all considered scenarios. Then, our objective function can be defined as:

$$MinF_1 = Min\{\frac{1}{N(N-1)}\sum_{i=1}^{N} y_{is}\}, \qquad \forall s \in S \qquad (2)$$

In equation (2), decision variable $y_{is}$ is 1 if node $i$ is not covered under scenario $s$, otherwise it is $0$; $i$, $I$ are the index and set of potential contamination source nodes, respectively; $N$ is the total number of nodes in the network; $s$, $S$ are the index and set of scenarios, respectively, and $N(N-1)$ is a constant which is used to normalize the objective function. If it is not utilized, the second function does not affect the solutions.

As second objective function $F_2$, we adopt a complex network metric, specifically the Average Node Coverage (ANC) introduced [52] as a new network diffusion speed metric describing how fast information (or entities) are distributed from a node (center) to other nodes (periphery). It computes the average number of nodes that receive information (or entities) in one time unit, given that all connected nodes receive information (or entities) and it takes one time unit for a node to pass information to all its directly connected nodes. Thus, $F_2$ is defined as:

$$MaxF_2 = Max \sum_{j=1}^{n} f_j (\frac{CS(z_j)-1}{LGD(z_j)}), \qquad \forall j \in J \quad (3)$$

In equation (3), decision variable $z_j$ is 1 if a sensor is located at node $j$ and $0$ otherwise; $j$, $J$ are the index and set of potential sensor location nodes, respectively; the component size, $CS(z_j)$, is the number of nodes in the component that contain node $z_j$; $CS(z_j)-1$ is the number of nodes that are connected to node $z_j$; $LGD(z_j)$ is the largest geodesic distance of the component which node $z_j$ belongs to; then,

$$\frac{CS(z_j)-1}{LGD(z_j)}$$

measures the average number of nodes that receive the information (or entities) originated from $z_j$, in one time unit. Because contamination diffusion speed in water distribution network through nodes that are connected to each other directly is important, this metric can be used to identify suitable sensor locations to detect contaminants in water distribution networks as early as possible. It is expected that the optimal solution of this objective function allows covering a maximum fraction of WDN nodes [52].

The criticality of nodes in a water distribution network is partially a function of the node type and associated hydraulic attributes, such as outflow, total demand, and hydraulic head. In this paper, for normalization of the second objective function $F_2$ we use the "node-adjusted entropic degree" measure [29]. This measure uses a dimensionless weighting factor to incorporate the nodal base demand for water in the definition of entropic degree. In other words, this factor adjusts the entropic degree by weighting it by the relative demand for water at each node [29]. Demand-adjusted entropic degree of node $i$ is defined as,

$$f_i = \frac{g_i}{2}(1+\frac{d_i}{M_d})$$

where $g_i$ is the entropic degree of node $i$, defined as,

$$g_i = (1-\sum_k p_{ik} \log p_{ik})\sum_k w_{ik}$$

and $d_i$ is the base (nominal) demand for water at node $i$ (in liters per second) and $M_d$ is the maximum of all such base demand values of the network nodes; $p_{ik}$ is the normalized weight of the link between nodes $i$ and $k$, with the property of $\sum_k p_{ik} = 1$ and defined as $p_{ik} = \frac{w_{ik}}{\sum_k w_{ik}}$, where $w_{ik}$ is the physical capacity of a pipe between nodes $i$ and $k$, defined as $w_{ik} = \frac{\pi L_{ik} D_{jk}^2}{4}$, where $L_{ik}$ is the pipe length and $D_{ik}$ is the pipe diameter between nodes $i$ and $k$ [29].

Thus, with equations (2) and (3) we have completed the building of our two objective functions. Note, that the coverage from a complex network perspective defined in (3) conflicts with the coverage considered in classical sense as defined in equation (2), as will be further shown in our case study of Section 4.

### 3.1.2. Constraints

We consider the following two constraints, originally proposed in Xu et al. [39]:

$$\sum_{j=1}^{n} z_j = p, \qquad \forall j \in J \qquad (4)$$

$$\sum_{j \in N_{is}} z_j + y_{is} \geq 1, \qquad \forall i \in I, s \in S \qquad (5)$$

Constraint (4) specifies the number of available sensors and $p$ is the total number of available sensors (for our calculations, we assume $p=5$). Constraint (5) stipulates that node $i$ is either covered by the sensors ($\sum_{j \in N_{is}} z_j \geq 1$) or uncovered ($y_{is} = 1$), under the possible scenarios of contamination. Since the first objective $F_1$ is to minimize the number of nodes uncovered, $y_{is}$ is forced to be *0* whenever $\sum_{j \in N_{is}} z_j \geq 1$. Under this constraint, $N_{is} = \{j: t_{ijs} \leq T\}$ is the set of candidate sensor-location nodes that can detect a contamination event occurring at node $i$ within time T under scenario $s$, $t_{ijs}$ is the contaminant travel time from contamination source node $i$ to potential sensor location node $j$ under scenario $s$ and $T$ is the ''coverage time'', i.e., the maximum allowable time from the start of the contamination event to detection. $z_j \in \{0,1\}$, $y_{is} \in \{0,1\}$ are integrality constraints [39].

### 3.2. Solution Method

For the solution of the bi-objective optimization problem, we adopt a weighting method whereby the objective functions are combined through a weight factor $W$:

$$MinF = WF_1 - (1-W)F_2 \qquad (6)$$

As the objectives are conflicting, we need to find the non-dominated or Pareto optimal frontier [51], by letting the value of $W$ change to produce some of the non-dominated solutions.

### 4. Case study

We consider a WDN comprise of 129 nodes, one constant head source, two tanks, 168 pipes, two pumps, eight valves (Figure 3). This WDN has been provided by the organizers of the Battle of Water Sensor Networks (BWSN) in a scientific competition held in 2006 [5]. Given a set of boundary conditions, such as four variable water demand patterns, we use the EPANET hydraulic simulator [53] to analyze network physical properties such as the steady-state pressure and obtain important output, such as the water flow velocity which is necessary to assign covering nodes. A sample of the results is shown in Table 3. We use $L_{ik}$ and $D_{jk}$ to estimate $w_{ik} = \pi L_{ik} D_{jk}^2 / 4$; then, $p_{ik} = w_{ik} / \sum_k w_{ik}$ and $g_i = (1 - \sum_k p_{ik} \log p_{ik}) \sum_k w_{ik}$ eventually $f_i = \dfrac{g_i}{2}(1 + \dfrac{d_i}{M_d})$.

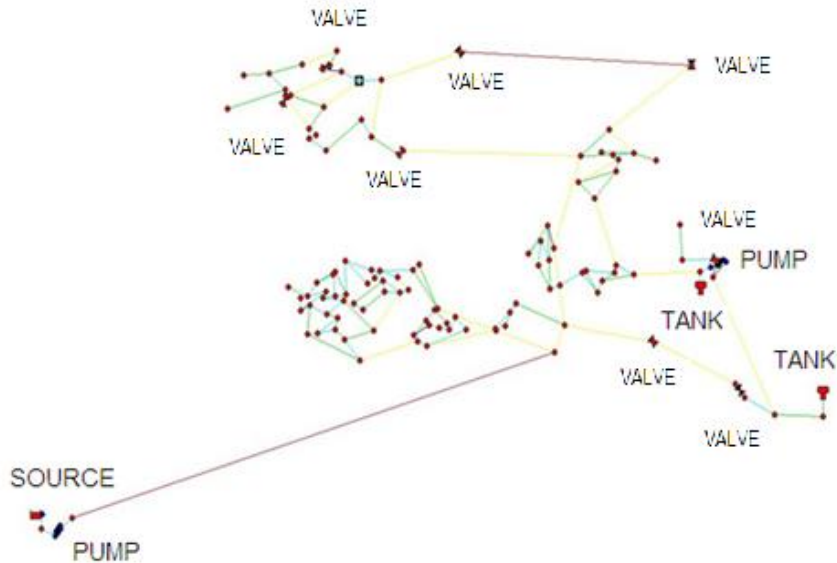

Figure 3. A WDN layout (BWSN) with 126 nodes, 1 source, 2 tanks, 168 pipes, 2 pumps, 8 valves [5]

Table 3. EPANET's sample results for the WDN of Figure 3

| Link ID | Start Node $i$ | End Node $k$ | Length (*ft*) $L_{ik}$ | Diameter (in) $D_{ik}$ | Velocity |
|---|---|---|---|---|---|
| LINK-0 | 118 | 126 | 7401 | 20.453411 | 0.3 |
| LINK-1 | 17 | 0 | 359 | 7.999894 | 0.32 |
| LINK-2 | 0 | 2 | 324 | 7.999894 | 0.16 |
| LINK-3 | 3 | 4 | 697 | 7.99981 | 0.09 |
| LINK-6 | 5 | 4 | 574 | 7.999807 | 0.3 |
| LINK-7 | 5 | 6 | 238 | 7.999912 | 0.49 |
| LINK-8 | 6 | 7 | 502 | 8 | 0 |
| LINK-9 | 8 | 6 | 396 | 7.999894 | 0.51 |
| LINK-10 | 9 | 8 | 377 | 7.999894 | 0.32 |
| LINK-19 | RESERVOIR-129 | 128 | 51 | 30 | 1.09 |
| LINK-72 | 29 | TANK-130 | 242 | 24 | 0.25 |
| LINK-165 | 102 | TANK-131 | 559 | 16 | 0.67 |
| PUMP-170 | 105 | 106 | #N/A | #N/A | 0 |
| PUMP-172 | 109 | 110 | #N/A | #N/A | 0 |
| VALVE-173 | 111 | 112 | #N/A | 6 | 0.03 |
| VALVE-180 | 125 | 126 | #N/A | 8 | 0 |

The scenarios considered for the network analysis are taken from [39]. Simulation of random contamination events was conducted with distinct contamination events beginning every 5 minutes for the first 24 hour, with each event lasting for 2 hours. A suite of $129 \times 24 \times 60/5$ contamination events was simulated for each of the 129 nodes in the network, yielding a total of 37,152 contamination events. In our analysis, we focus on accidental random contamination events rather than one intentional contamination event. We assume that the contamination event might occur at any node. For our purposes, one scenario is the injection at any of the nodes of the same amount of contaminant, at the same starting time and for the same duration. Thus, we have 24 × 60/5=288 scenarios, with each scenario starting every five minutes for the first 24 hours of the simulation [5].

Table 4. Sample computation of node weights in undirected WDN (Figure 3)

| Node ID | Nodal Demand($d_i$) | $g_i$ | $f_i$ | Normalized $f_i$ |
|---|---|---|---|---|
| JUNCTION-0 | 0.763534 | 153,730.59 | 77,162.21 | 0.01537 |
| JUNCTION-1 | 0 | 286,810.25 | 143,405.13 | 0.02857 |
| JUNCTION-2 | 0.83 | 94,329.38 | 47,362.74 | 0.00943 |
| JUNCTION-3 | 4.4 | 121,948.61 | 62,331.59 | 0.01242 |
| JUNCTION-4 | 3.836466 | 204,539.49 | 104,254.70 | 0.02077 |
| JUNCTION-5 | 3.529833 | 194,080.68 | 98,773.26 | 0.01968 |
| JUNCTION-6 | 1.85 | 83,280.85 | 42,030.15 | 0.00837 |
| JUNCTION-7 | 0 | 25,233.27 | 12,616.64 | 0.00251 |
| JUNCTION-8 | 1.54 | 50,545.37 | 25,469.58 | 0.00507 |
| JUNCTION-9 | 0 | 44,719.98 | 22,359.99 | 0.00445 |
| JUNCTION-10 | 0.341497 | 65,524.77 | 32,818.99 | 0.00654 |
| JUNCTION-129 | Reservoir | 36,049.78 | 1.00 | 0.00000 |
| JUNCTION-130 | Tank | 109,478.22 | 1.00 | 0.00000 |
| JUNCTION-131 | Tank | 112,393.62 | 1.00 | 0.00000 |

## 5. Results and discussions

In this section, we show how a maximum of $p=5$ sensors should be optimally placed in the 129-node network to obtain maximum coverage under 288 contamination scenarios. General Algebraic Modeling System (GAMS) [54] was used to formulate and solve the optimization. In Table 4, we have provided a sample computation of node weights in undirected WDN.

Note that $f_i = \dfrac{g_i}{2}(1+\dfrac{d_i}{M_d}) = \dfrac{153,730.59}{2}(1+\dfrac{0.763534}{197.663527}) = 77,162.21$

The normalized value is estimated, then, as 77,162.21/5020313=0.01537, as indicated in Table 4 at row 2.

In Table 5, we reveal the weighted objective function value $F$, based on different weight values. In this Table, "$W$" shows the range of weight under which the non-dominated solution does not change. For example for $W=0.6$ up to $W=1.0$, the weighted function "$F$" is equal to 1.500. The "$F1$" and "$F2$" values of the first and second objective functions, in this case, are equal to $F1=1.500$, $F2=-0.173$, respectively. Then, "sensor location node ID" values for this weighting range are 6, 28, 33, 34 and 58.

Table 5. Sensor nodes locations for different ranges of weight values

| Weight, $W$ | $F$ | $F_1$ | $F_2$ | sensor location node ID | | | | |
|---|---|---|---|---|---|---|---|---|
| 0.6-1 | 1.500 | 1.500 | -0.173 | 6 | 28 | 33 | 34 | 58 |
| 0.4-0.55 | 0.642 | 1.709 | -0.425 | 20 | 28 | 33 | 34 | 118 |
| 0.1-0.35 | 0.313 | 1.971 | -0.580 | 20 | 28 | 33 | 34 | 118 |
| 0-0.05 | -0.453 | 1.971 | -0.580 | 19 | 20 | 23 | 34 | 118 |

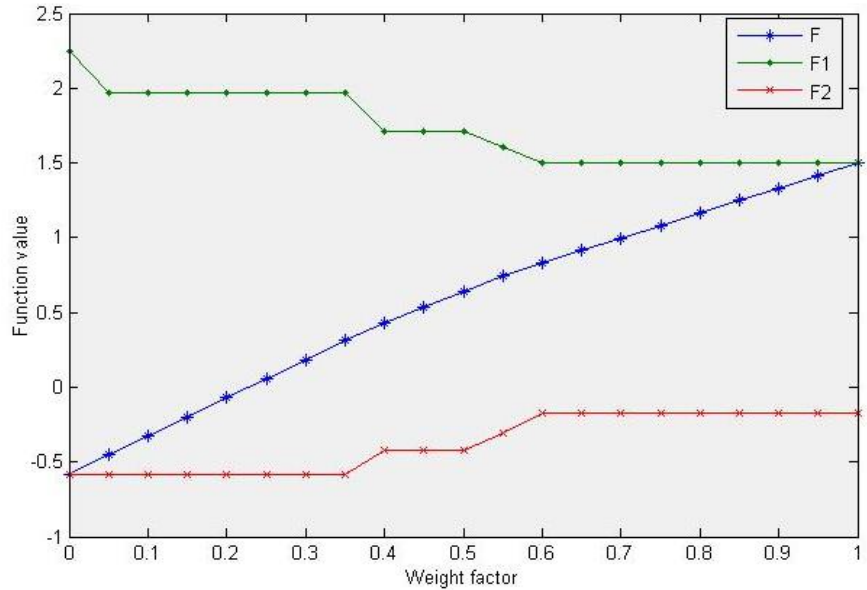

Figure 4. Objective function values $F$, $F_1$ and $F_2$ versus weight changes.

However, by changing the $W$ values from 0.6 to 0.55, the solution changes as shown in the third row of Table 5. Figure 4 shows the changing values of the objective functions $F$, $F_1$ and $F_2$ for the weights assigned to every objective. Overall, by changing the weight values, we have found three different Pareto optimal solutions of sensor location. It might be possible to find more solutions by introducing finer resolution on $W$ or adopting a different method such as the ε-constraint method [51]. Still, the fact that most nodes in the current solutions are common including 28, 33, 34 and 118 may indicates that we have obtained a rather stable solution set.

We pick the most frequent sensor node locations from Table 5, which is 20, 28, 33, 34 and 118, shown on the WDN topology of the BWSN in Figure 5. In this Figure, we have also shown the results for the single objective functions of $F_1$ and $F_2$, respectively optimized.

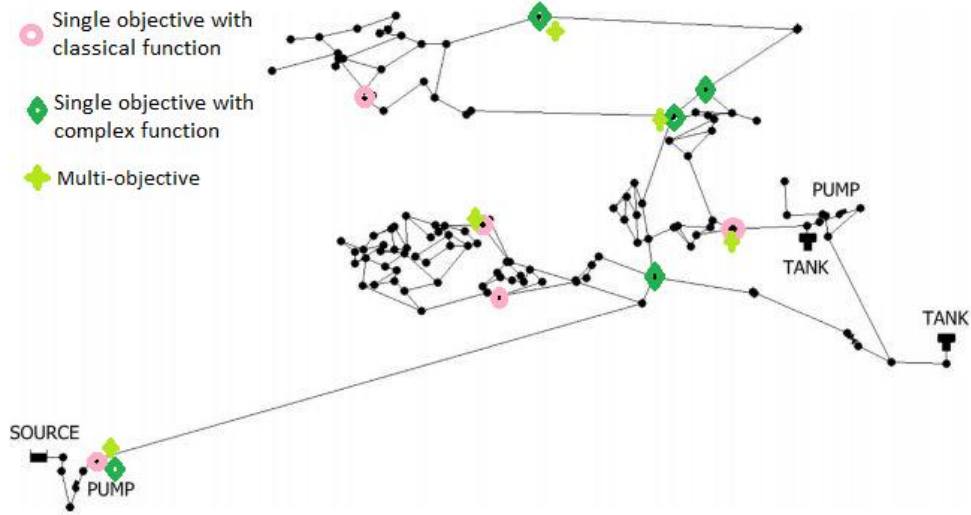

Figure 5. Optimal sensor node locations in the WDN network (BWSN), for the three cases of single-objective $F_1$, $F_2$ and bi-objective $F$ optimization

If we divide the network into two layers calling them the inner side (the core) and the outer side (the periphery), then the results coming from the complex theory objective function $F_2$ are distributed mostly in the inner side of the network, i.e., 4 out of 5 sensors. On the contrary, if the classical objective function $F_1$ is used, then 3 out of 5 sensors are placed on the outer side of the network. However, if bi-objective optimization is used, the sensor nodes are allocated both on the inner and outer sides.

Both objectives functions $F_1$ and $F_2$ are aimed to cover as much as nodes as possible, although from a different point of view, thus targeting different layers of the network. Instead, the Pareto optimal sensor location by the bi-objective model includes both objective functions.

Now, we turn our attention to adopt some of the metrics of complex network theory as described in Table 1 to define the distribution of sensor nodes. The nodes are selected based on the maximization of the degree, betweenness and closeness centrality [13], [14]. By considering three metrics we obtain the results reported in Table 6. For comparisons, we have also evaluated the values of $F_1$, $F_2$ and $F$, also reported in Table 6.

Table 6. Sensor nodes locations distribution by using the complex network metrics

| Metrics | $F$ | $F1$ | $F2$ | sensor location node ID | | | | |
|---|---|---|---|---|---|---|---|---|
| Degree | -0.061 | 2.880 | -0.061 | 20 | 22 | 35 | 92 | 97 |
| Betweenness | -0.277 | 2.880 | -0.277 | 20 | 22 | 23 | 30 | 31 |
| Closeness | -0.056 | 2.880 | -0.056 | 21 | 22 | 23 | 24 | 30 |

As shown in Figure 6, we notice again that most sensors are located on the inner side of the network.

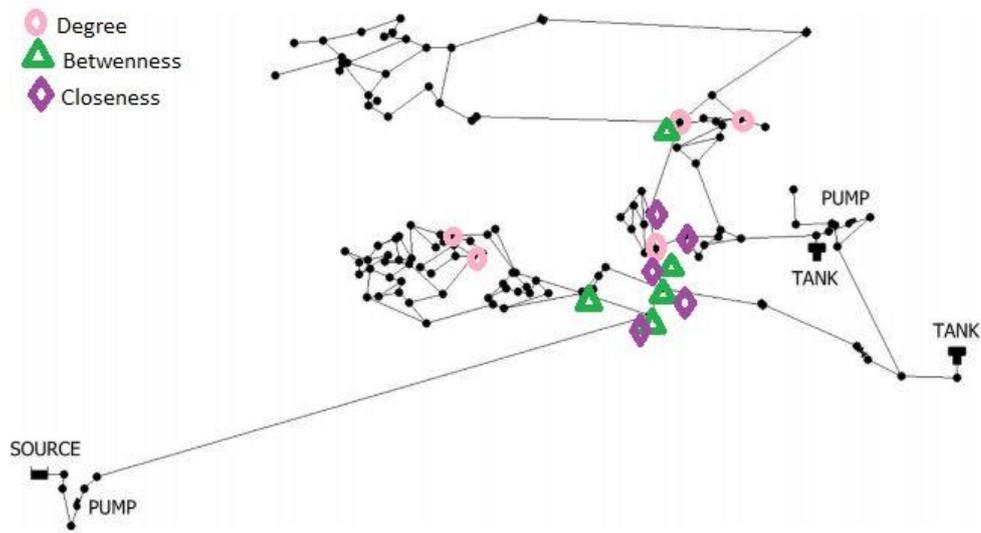

Figure 6. Distribution of sensor node locations in a WDN network (BWSN) based on the maximization of degree, betweenness and closeness centrality

We also compare our results with those obtained in 15 other works on the BWSN network, as shown in Figure 7. In this Figure, we show the sensor nodes locations obtained in the scientific competition (BWSN) in 2006 and reported in [5]. On the same Figure 7, we have interested two solution sets, namely one from Xu et al. [39] and the results obtained by our bi-objective optimization.

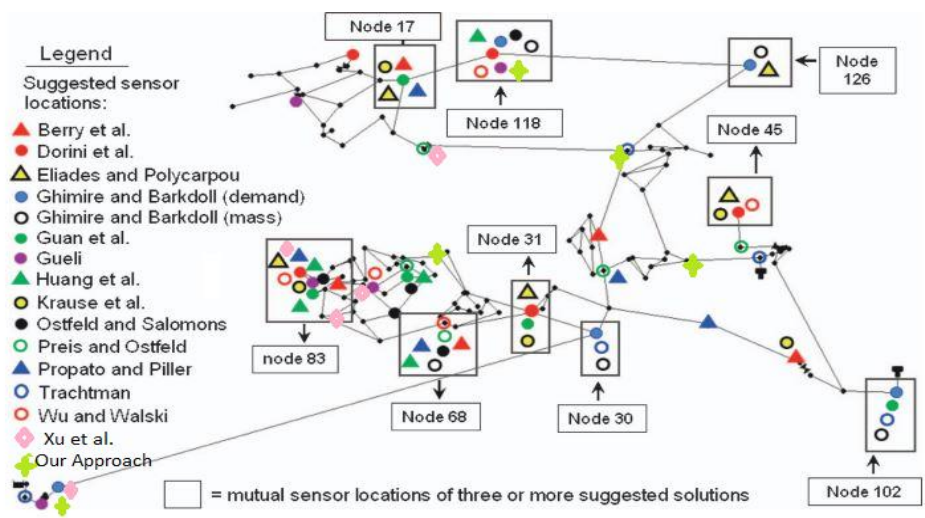

Figure 7. WDN network (BWSN) and alternative sensor nodes location solutions.

Xu et al. [39] have adopted three different models of minimax, minimax regret and deterministic model and produced four different solution sets given in Table 7.

Table 7. Solution set obtained by Zu et al. [39]

| Approach | Solution Set( sensor location node ID) |
|---|---|
| Minimax approach | a. [72,83,100,116,121] <br> b. [74,83,100,116,121] |
| Minimax regret approach | [34,74,83,100,116] |
| Determinstic model | [74,83,100,120,124] |

### 5.1. Performance of the new approach

Rathi and Gupta [4] in their review in 2014 have mentioned that " it can be observed from the reported literature that there is no consensus amongst researchers on the number and type of performance objectives to be considered and several other issues related to sensor location problem" (page 186). Results reported in Figure 7 and Table 7 also reinforce the lack of agreements on how the performance of a sensor placement should be measured. We consider the following two principles for performance measurement of laying out the sensors and based on them we show that the modeling approach developed in this paper performs better than the other models.

**First principle:** Since a WDN is a geographically distributed network, so should be the sensors. By looking at Figure 7, we see that the 65 sensors locations proposed by 15 previous works (each one considering a maximum of 5 sensors) are not uniformly distributed. For example, around node 83 we see that 13 sensors are assigned, whereas at node 102 or 45 only a single sensor is assigned. In Figure 8, we have reproduced Figure 7 but added the five circles of the sensors which have been located by the new modeling approach. Though, the sensors are not located at the center of the circles, they cover almost all the geographical area of the WDN.

**Second principle:** Since a WDN is a complex network, any sensor placement method should consider this characteristic. The sensor placements in Figure 6, which are based on pure complex network metrics like degree, betweenness and closeness centrality, favor inner side nodes. We tested this property once more in Figure 5, where we considered the complex-based objective function $F_2$. Again, when the complex-based measure is used, the sensors are mostly located in the inner side of the WDN, whereas by using the classical model they are mostly located at the periphery of the WDN. Finally, the results produced by our bi-objective optimization accommodate both principles outlined above.

In addition, the WDN considered here has only one water source. It can be argued that a viable sensor placement solution should assign one sensor to this source node as it is the most important node for intentional (or terrorist) attacks: any contamination inflow here in the source node will distribute to the other nodes in the network. In our solution, one sensor is, indeed, reasonably, assigned to the source node, whereas it can be seen from Figure 7 that only 4 out of 15 models did so.

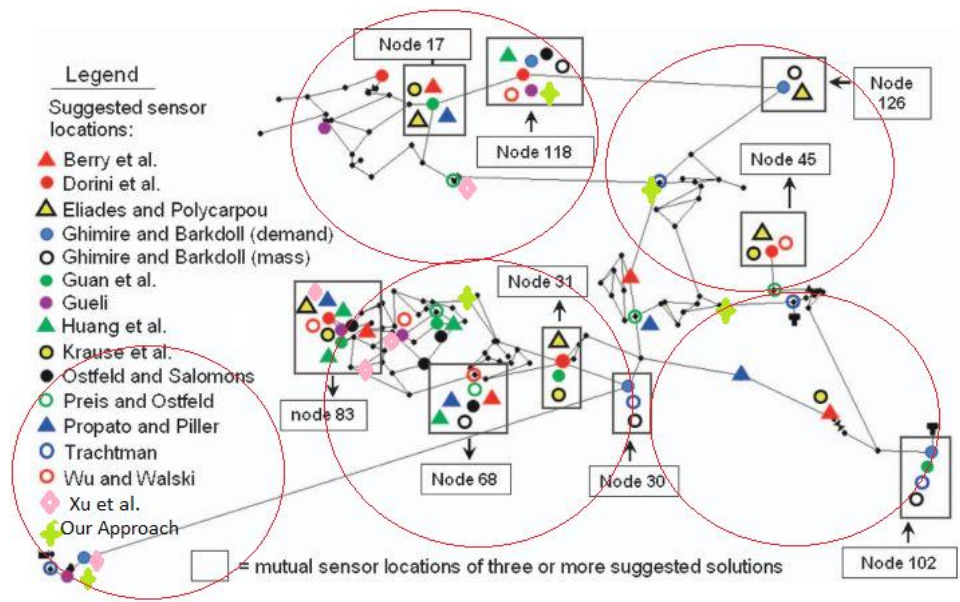

Figure 8. Five circles covering the WDN

## 6. Conclusions

The goal of sensor placement in WDNs is to detect contamination of intentional and accidental events, and to ensure safe and quality drinking water for consumers. Solving this problem as a classical optimization problem has been the focus of attention for many years. Models with single and multiple objectives have been formulated, and techniques such as integer and mixed-integer programming, stochastic programming, robust optimization, branch and bound, heuristics and meta-heuristics such as greedy and genetic algorithms have been applied. The original contribution of this paper is to present a new modeling approach by taking into consideration the fact that a WDN is a complex network. Thus, the structural and dynamical complexity properties should play a role in our sensor placement problem. Hence, we model the sensor placement problem within the complex network theory paradigm, and propose an integrated approach using both the classical and complex network perspective.

For practical reference, we have considered a well-known water network, with 129 nodes and 178 links to show the working logic of our new approach. The test bed we adopted is a WDN previously considered for the sensor location problem, by about 20 other research groups. This was helpful to present a meaningful platform for comparison of the results across a wide range of approaches adopted by researchers in the field.

The advantages of the new integrated approach proposed have turned out to be: 1) likewise the literature, we have formulated the sensor placement problem as a multi-objective optimization problem; 2) unlike in the literature, our approach is based on the fact that a WDN is a complex network; 3) our results reasonably match the results of others; 4) still our results improve the results of others, from the view point of complex network properties.

The next step in the research is to formulate and solve a sensor placement problem in WDNs as a many-objective optimization problem, since there is no consensus amongst researchers on what objectives should be considered, like minimizing the detection time, the volume of contaminated water, the exposed population to contamination, the number of failed detections, the probability of failed detection, the sensor response time, the number of sensors, the discrepancy between residential and industrial consumption patterns, the discrepancy between time patterns, the discrepancy between intentional and accidental attacks. Also, it is equally important to maximize the coverage, the sensor detection likelihood and the sensor detection redundancy. How to consider many of these in an effective optimization framework and how to eventually take final location solutions is an interesting issue to address.

A second line of further research is to include the complex network properties of a WDN into the structure of the constraints of the optimization problem, thus leading to a full integration of the complex network theory and the many-objective optimization theory.

**Acknowledgment**

Comments made by Dr. Alireza Yazdani when he was post-doctoral research associate at the department of Civil & Environmental Engineering in Rice University on the early stages of development of this research have been very helpful and encouraging. We are grateful.